# Paramagnetic resonance in La$_2$NiMnO$_6$ probed by impedance and lock-in detection techniques

Ushnish Chaudhuri[1], Debendra Prasad Panda[2], A. Sundaresan[2] and R. Mahendiran[1]

[1]Physics Department, 2 Science Drive 3, National University of Singapore, Singapore-117551, Republic of Singapore

[2]Chemistry and Physics of Materials Unit, Jawaharlal Nehru Centre for Advanced Scientific Research, Jakkur P.O., Bangalore-560064, India

**Abstract**

We report the detection of paramagnetic resonance in the double perovskite La$_2$NiMnO$_6$ at room temperature for microwave magnetic fields with frequencies, $f$ = 1 GHz to 5 GHz, using two cavity-less methods. We use an indirect impedance method which makes use of a radio frequency impedance analyzer and a folded copper strip coil for the frequency range $f$ = 1 to 2.2 GHz. In this method, when an applied *dc* magnetic field is swept, high-frequency resistance of the strip coil exhibits a sharp peak and the reactance curve crosses zero exhibiting resonance. A lock-in based broadband setup using a coplanar waveguide for microwave excitation was used for $f$ = 2 to 5 GHz The resonance fields ($H_r$) obtained from both the techniques increase linearly with frequency and a large spectroscopic g-factor, equal to 2.1284, which supports the presence of Ni$^{2+}$ cation with strong spin-orbit coupling. Line shape analysis and analytical fitting were performed to characterize the material in terms of its initial susceptibility and damping parameters.

[1] Author for correspondence(phyrm@nus.edu.sg)

La$_2$NiMnO$_6$ (LNMO) belonging to the double perovskite family A$_2$BB′O$_6$ is a ferromagnetic insulator with a Curie temperature close to room temperature (T$_C$ = 270 K). This compound has been a subject of intensive studies in recent years because of its magnetocapacitance effect (modulation of its dielectric constant by external magnetic field).[1,2,3,4,5] Ferromagnetism in LNMO is due to superexchange interaction between half-filled e$_g$ orbitals of Ni$^{2+}$ (t$_{2g}^6$e$_g^2$, S=1) and empty e$_g$ orbital of Mn$^{4+}$ (t$_{2g}^3$e$_g^0$, S=3/2) ions.[6,7] Saturation magnetization and $T_C$ attain maximum values when Ni$^{2+}$ and Mn$^{4+}$ ions are ordered alternatively at B and B′ sites. Anti-site defects i.e., presence of Ni$^{2+}$ on Mn$^{4+}$ site and vice versa weakens ferromagnetism and promotes antiferromagnetic interactions leading to spin glass or cluster glass behavior at low temperatures.[8,9] While several studies have focused on static magnetic properties, dynamic magnetic properties of LNMO received less attention so far. Based on *dc* magnetization and line shape analysis of electron spin resonance (ESR) spectra, Zhou *et al.*[10] suggest ferromagnetic correlations persisting up to 90 K above T$_C$ in bulk LNMO samples. From angular dependence ESR studies, Kazan *et al.*[11] could detect a signature of ferromagnetic resonance (FMR) at room temperature in LNMO films grown on (100) oriented SrTiO$_3$ and (1110) oriented NdGaO$_3$. They suggest coexistence of two magnetic phases with different easy and hard axis periodicities in the plane of LNMO. Mallick *et al.*[12] reported spin Hall *dc* magnetoresistance in LNMO/Pt bilayer while Shimoi *et al.*[13] reported paramagnetic spin pumping in LNMO/Pt bilayer films. Pure spin current during paramagnetic resonance was pumped from LNMO into the Pt layer and detected via the inverse Hall effect. While it is established that ferromagnet exposed to continuous wave microwave magnetic field can pump pure spin current into an adjacent heavy metal during ferromagnetic resonance, spin pumping in the paramagnetic state of LNMO is surprising. These studies underscore the importance and necessity of dynamic magnetic measurements.

All the available ESR studies on LMNO were done at a single frequency (f ~ 9.8 GHz) of microwave irradiation and the resonance occurred around 3.4 kOe. Here, we report magnetic resonance at lower frequencies and hence at smaller magnetic fields using two different cavity less methods: An impedance based method using a copper stripcoil and an *rf* impedance analyzer was used for the frequency range $f$ = 1 MHz to 2.2 GHz, and a lock-in amplifier based broadband FMR spectrometer setup using a coplanar waveguide (CPW) was used in the frequency range $f$ = 2 to 5 GHz, The gyromagnetic ratio, $\frac{\gamma}{2\pi} = \frac{g\mu_B}{\hbar}$ was obtained by fitting the resonance field linearly with excitation frequency and the spectroscopic '$g$' factor was obtained. A broadband approach can accurately measure the $g$ of LMNO, which has remained a contentious topic so far. Zhou *et al.*[11] reported $g$ ~2.0, while Shiomi *et al.*[14] stated $g$ ~2.14.

Recently, we have shown that paramagnetic resonance could be excited in the manganite $La_{0.6}Ca_{0.4}MnO_3$ without using a microwave cavity or other transmission techniques such as coplanar waveguide and microstrip.[14] The conducting sample itself acted as a waveguide in presence of radio frequency (*rf*) current in the sample. Upon sweeping the *dc* magnetic field that was applied parallel to the direction of rf current and hence perpendicular to *rf* magnetic field, high frequency resistance showed a sharp peak at a critical value of magnetic field. As LNMO is highly resistive (~ 70 MΩ at 300 K), *rf* current could not be passed directly through LNMO. Hence, we adapt an alternative method which makes use of a copper strip coil and an impedance analyzer.[15] The impedance analyzer measures resistance and reactance of the copper strip enclosing the LNMO sample using the current-voltage method at several frequencies ($f$ = 1 MHz to 3 GHz) of *rf* current. The measured impedance of the copper strip reflects changes in the dynamic magnetic properties of the sample as dc magnetic field is varied. Since natural resonance of the strip coil was around 2.6 GHz, another spectrometer was used to extend the frequency range and verify the results obtained using the impedance technique. Our lock-in based broadband resonance spectrometer makes use of a microwave source (Anirtsu MG3696) to inject *rf* signal into a 50 Ω impedance matched CPW and an electromagnet to vary the *dc* magnetic field. A lock-in amplifier (SRS830) is used to modulate the *dc* magnetic field at a frequency of 197.4 Hz with the help of a pair of Helmholtz coils, the *rf* signal picked up in the CPW is rectified by a Schottky zero bias detector (PE8013) and is fed to the lock-in amplifier. The working principle of this technique is discussed in recent papers.[16] The *dc* output of the lock-in amplifier is proportional to d$P$/d$H$ where $P$ is the microwave power absorbed by the sample and $H$ is the dc magnetic field. The LNMO sample was prepared by solid-state reaction method and characterized by X-ray diffraction. The sample was found to be single phase and crystalized in monoclinic structure (space group P2$_1$/n).

The main panel of Fig. 1 shows magnetization ($M$) versus temperature data recorded while cooling the sample from 400 K to 100 K in a magnetic field of $H$ = 1 kOe. The ferromagnetic temperature is $T_C$ = 271.6K as obtained from the inflection point of d$M$/d$T$ curve. The inverse susceptibility shown on the right-y axis is fitted with the Curie-Weiss linear law ($H/M = C/(T-\theta_p)$) above 320 K and shows a small deviation from the linear fit below 320 K. From the Curie-Weiss fit, we obtain the paramagnetic Curie temperature $\theta_p$ = 279.6 K and the effective magnetic moment P$_{eff}$ = 5.357 µ$_B$ which is smaller than the theoretical estimate P$_{eff}$ = 6.701 µ$_B$. The smaller effective moment value indicates the presence of weak antiferromagnetic interactions in the system due to antisite defects. The $M$ vs $H$ curve at 300 K (inset) shows linear variation $M$ with the magnetic field up to 10 kOe reflecting the paramagnetic behavior.

In Fig. 2 (a) and (b) we present the 3-D plots of the high-frequency resistance ($R$) and reactance ($X$) of the copper strip coil enclosing the sample as a function of *dc* magnetic field ($H$) for several frequencies of the *rf* magnetic field. *R* and *X* values for each frequency are subtracted from their respective values for the empty strip coil. Insets to the top and bottom graphs illustrate the field dependence of *R* and *X* at $f = 1$ GHz in an enlarged scale. As *H* is decreased from the maximum value, *R*(*H*) is nearly field independent down to 700 Oe, below which it increases rapidly to exhibit a peak at 356.3 Oe where $R = 0.032$ $\Omega$ and decreases at lower fields. The main panel indicates that resistance peak increases in magnitude and its position shifts towards higher magnetic fields with increasing frequency. The line shape is that of a Lorentzian. The peak value of resistance increases by hundred folds to 7.265 $\Omega$ at $H = 735.2$ Oe when $f = 2.2$ GHz. In contrast to *R*(*H*), *X*(*H*) shows dispersive behavior and crosses over zero value at the field value where *R*(*H*) exhibits a peak. Both *R*(*H*) and *X*(*H*) exhibit mirror symmetry about the origin upon reversing the field direction. The observed excess resistance of the strip coil is caused by a magnetic loss in the sample, related to the out of phase component ($\chi''$) of the high frequency complex susceptibility ($\chi = \chi' - i\chi''$). The field dependence $\chi''$ exhibits a peak and $\chi'$ sharply decreases and crosses zero when resonance occurs in a spin system. The reactance is proportional to $\chi'$ and hence shows a change of sign during the resonance. In contrast to the present result, both absorptive ($\chi''$) and dispersive ($\chi'$) signals were mixed up in the resistive and reactance components of less resistive $La_{0.6}Ca_{0.4}MnO_3$ sample because of the skin effect.[16]

Fig. 3(a) shows the magnetic field dependence of the derivative of the power absorption (d*P*/d*H*) obtained from the lock-in FMR setup for $f = 2$ GHz to 5 GHz in steps of 0.2 GHz. The field was swept from +2 kOe to -2 kOe. Hysteresis was not found upon reversing the field direction. We can notice the resonance feature of a negative peak followed by a positive peak as like in a cavity based ESR method. However, the curve does not display mirror symmetry about the origin, instead shows dP(-H)/dH = -dP(H)/dH. The resonance feature shifts towards higher field with increasing frequency, similar to the trend seen in the impedance measurement. In Fig. 3(b) we show $R(H)$, $X(H)$ obtained from the impedance measurements along with the derivative $dR/dH$ and $dP/dH$ from the lock-in measurement for comparison. We can see that $dR/dH$ curve resembles the $dP/dH$ curve indicating similar origins in both measurements. The high frequency $R(H)$ curve contains information about the power loss in the sample.

To extract the resonance field ($H_r$) and the linewidth ($\Delta H$), the impedance data was fit to a linear combination of symmetric and antisymmetric Lorentzian functions.

$$R \text{ or } X = K_{sym} \frac{(\Delta H)^2}{(H - H_r)^2 + (\Delta H)^2} + K_{asym} \frac{(\Delta H)(H - H_r)}{(H - H_r)^2 + (\Delta H)^2} + C \quad (1)$$

While, the lock-in data was fit to the derivate of Eq. 1, shown in Eq.2 after simplification.

$$\frac{dP}{dH} = A_{asym} \frac{(\Delta H/2)(H - H_r)}{[(H - H_r)^2 + (\Delta H/2)^2]^2} - A_{sym} \frac{(\Delta H/2)^2 - (H - H_r)^2}{[(H - H_r)^2 + (\Delta H/2)^2]^2} + B \quad (2)$$

$K_{sym}$, $A_{sym}$ and $K_{asym}$, $A_{asym}$ are the frequency dependent magnitudes of the symmetric and asymmetric dispersive components of the signals respectively. They depend on the phase between the *rf* electric and magnetic fields inside the material. $B$ is the offset parameter. The fit is shown as solid black lines for both $R(H)$, $X(H)$, and $dP/dH$ in Fig. 3(b) and closely reproduces the obtained results.

Fig. 4(a) shows the frequency dependence of $H_r$. The $H_r$ obtained from the impedance and the lock-in measurements fall on a single curve and linearly increases with the frequency. The data is fitted to resonance condition $H_r = 2\pi f/\gamma$ which gives the slope $2\pi/\gamma = 0.3356$ kOe/GHz or $\gamma/2\pi = 2.979$ GHz/kOe. This value of is larger than free electron value of $\gamma/2\pi = 2.8024$ GHz/kOe assuming spin only contribution. Using the experimental $\gamma/2\pi$ value, $R$ and $X$ were analytically fitted to the field dependent impedance at 2 GHz as shown in Fig. 4 (b). $R$ and $X$ components of the complex electrical impedance ($Z$) of the copper strip coil depend on the relative permeability ($\mu_r$) of the sample through the relations $R = K\sqrt{(\omega\mu_0\mu_r'')}$ and $X = K\sqrt{(\omega\mu_0\mu_r')}$, where, $\omega = 2\pi f$ is the angular frequency, $\mu_o$ is the free space permeability, $\mu_r'$ and $\mu_r''$ are the real and imaginary components of relative permeability of the sample and $K$ is a constant related to the geometry of the strip coil. Thus, changes in the permeability of the sample in response to variations of magnetic field and frequency give rise to changes in impedance. The dynamical permeability ($\mu = \mu' - j\mu''$) obtained using the Landau-Lifshitz-Gilbert (LLG) equation is given by:[17]

$$\mu(H) = 1 + \frac{\left[\chi_0 \left(1 + i\alpha\left(\frac{f}{f_r}\right)\right)\right]}{\left[\left(1 + i\alpha\left(\frac{f}{f_r(H)}\right)\right)^2 - \left(\frac{f}{f_r(H)}\right)^2\right]} \tag{3}$$

where resonance frequency $f_r$ is dependent on magnetic field ($H$) and $\alpha$ is the unit less Gilbert damping parameter. The linewidth is $\Delta H = 4\pi\alpha f/\gamma$ for ferromagnetic resonance and $\Delta H = 4\pi f/H\gamma^2\tau$ for paramagnetic resonance, where $\tau$ is the Bloembergen-Bloch relaxation time (spin-spin relaxation time). The Gilbert damping parameter is related to $\tau$ by $\alpha = 1/\tau\gamma H$.[18] The analytical fit with $\gamma/2\pi = 2.979$ was used to obtain the key parameter values of $\alpha = 0.14$ and the initial susceptibility, $\chi_0 = 0.43$. We have used $\alpha = 0.14$ to plot the variation of $\Delta H$ with frequency and it agrees well with the $\Delta H$ value obtained from the line shape analysis, which can be seen in the inset of Fig. 4(a). $\Delta H$ at 2 GHz was around 79.4 Oe while at 5 GHz was around 111.69 Oe.

The large value of $\gamma/2\pi = 2.979$ GHz/kOe implies a g-factor of 2.1284 which is far away from 2.0023. This is in stark contrast with the $Mn^{3+}$-$Mn^{4+}$ pair in manganites for which the g value is close to 2 in the paramagnetic state.[19] We obtained a g-factor of 1.999 for the standard EPR marker DPPH using the same magnetoimpedance setup and hence the results are reliable.[20] The large g-factor in LNMO indicates significant contribution of orbital magnetic moment and hence a strong spin-orbit coupling in $Ni^{2+}$ cation. This is also in agreement with an X-ray magnetic circular dichroism results which suggests that ratio of orbital to spin moment for $Ni^{2+}$ is an order of magnitude higher than for $Mn^{4+}$ cation.[21] Also, the presence of super exchange mechanism via virtual exchange of exchange of $e_g$ electron in $Ni^{2+}$-$Mn^{4+}$ gives rise to intense ESR signal as was the case for doped manganites whereas $e_g$ electron hopping between Mn3+ and Mn4+ ion leads to intense, narrow ESR signal in the paramagnetic state. The $Mn^{4+}:t_{2g}^3$ (S= 3/2) ion in octahedral oxygen environment leads to weak ESR signal at room temperature with g ~ 2 in lightly La-doped $CaMnO_3$.[22] The presence of $Ni^{2+}$ ions in the local vicinity of $Mn^{4+}$ can enhance the ESR signal as observed in the $LiCo_{1-2x}Ni_xMn_xO_2$ samples[23] in which an increase in the g-value with $x$ reflected the presence of more $Ni^{2+}$ ions around $Mn^{4+}$. It will be interesting to investigate how the g-value changes in $LaNi_{1-x}Mn_xO_3$ series in which the charge on the Ni cation shifts from 3+ for $x = 0$ to 2+ for $x = 0.5$.[7]

In summary, we have investigated the change in impedance as well as the change in the derivative of the power absorbed by a paramagnetic LNMO sample, while varying *dc* magnetic fields for different excitation frequencies. ESR features were observed and analyzed in terms of their line shape and a large g-factor was obtained from the ESR data which reflected the influence of $Ni^{2+}$ ions in the sample. The analytical fit also agreed well with the observed data.

**Acknowledgments:** R.M acknowledges the Ministry of Education, Singapore for supporting this work (Grant numbers: R144-000-381-112 and R144-000-404-114). A.S acknowledges the International Centre for Materials Science, JNCASR, for experimental facility.

**Data Availability Statement**: The data that support the findings of this study are available from the corresponding author upon reasonable request..

**Figure Captions:**

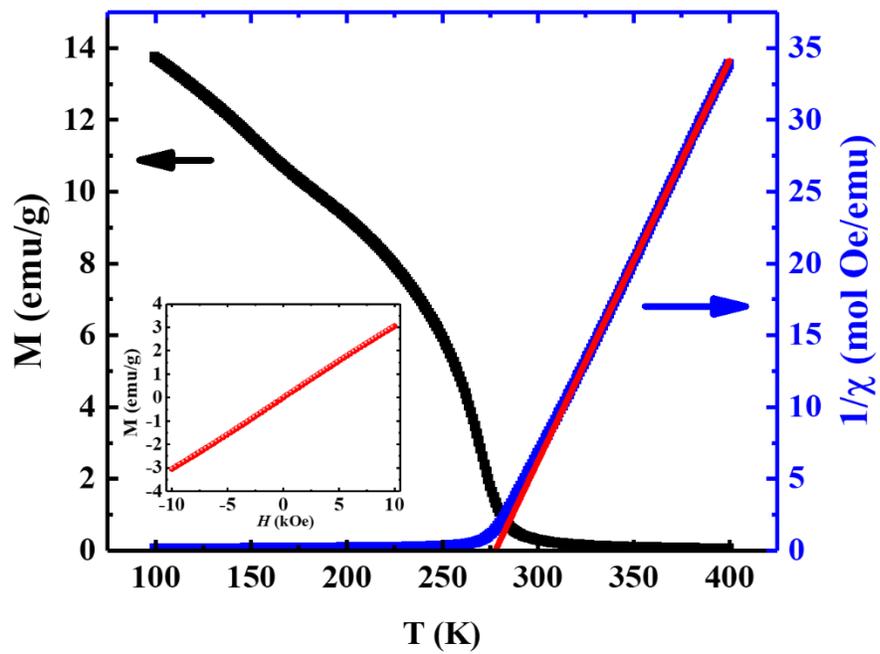

**Fig. 1.** (a) M-T (left scale) and $1/\chi$ (right) curves for $La_2NiMnO_6$. Solid line represents the Curie Weiss linear relationship. Insert: M-H at 300K

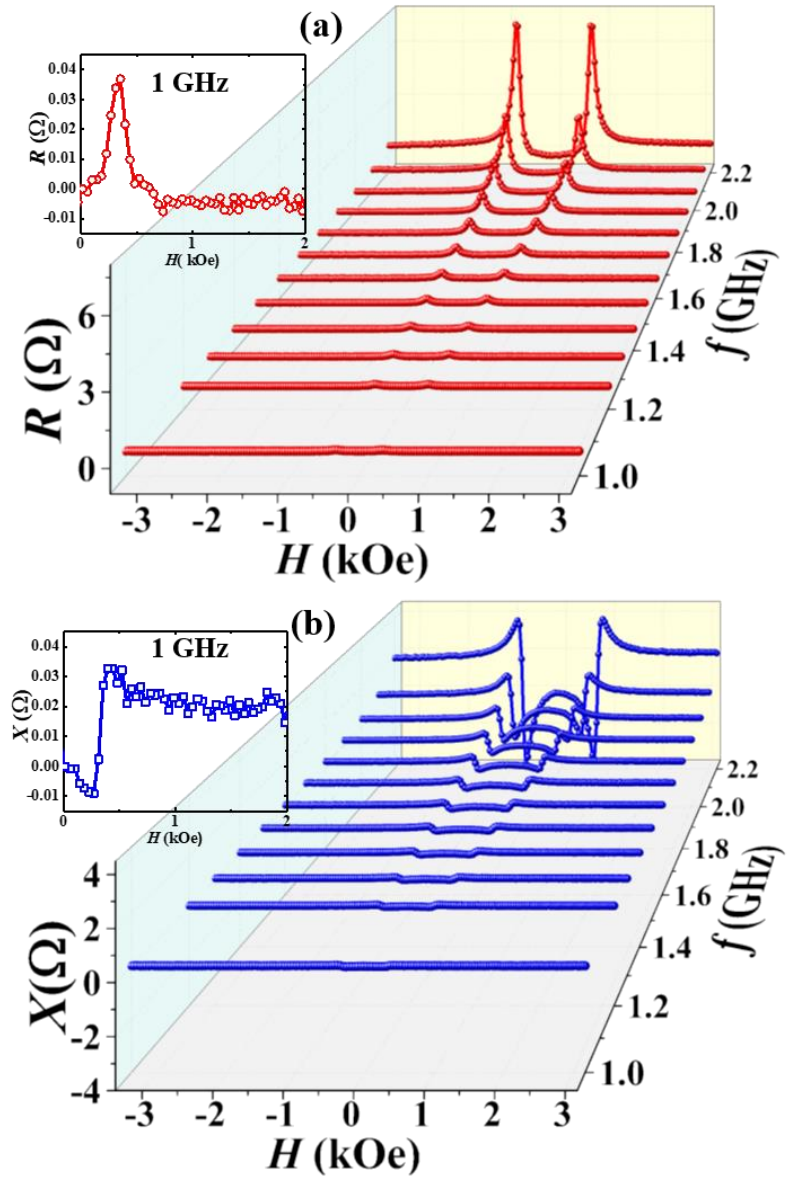

**Fig. 2** Magnetic field dependence of high frequency (a) resistance (*R*) and (b) reactance (*X*) for varying frequencies current in the copper stripcoil. Top and bottom insets represent *R* and *X* measured at 1 GHz.

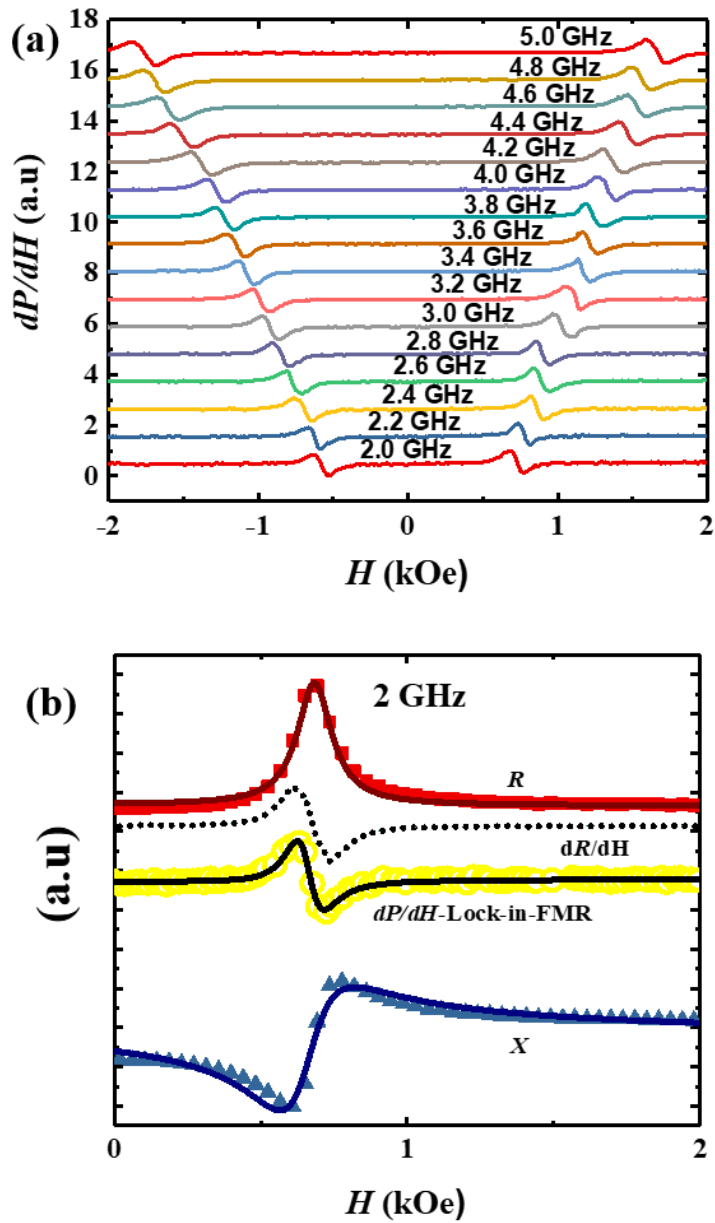

**Fig. 3.** (a) d*P*/d*H* response for varying excitation frequencies as measured by the Lock-in. (b) R, X, d*P*/d*H* and d*R*/d*H* measured at 2GHz. Solid lines represent the line shape fitting using Eq.1 and 2.

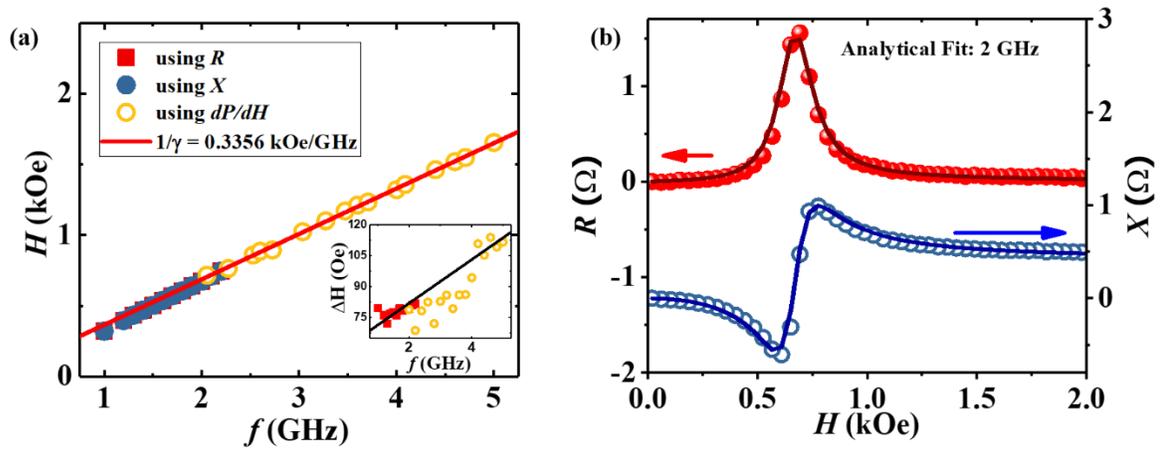

**Fig. 4.** (a) Resonance fields ($H_r$) obtained corresponding to their frequency of excitation using the line shape fits for *R, X,* and *dP/dH*. Solid line is the linear fit according to the equation: $H_r = 2\pi f/\gamma$. Inset: The linewidth $\Delta H$ obtained from the line shape analysis. Solid line represents the linear relationship of $\Delta H$ vs $f$ with Gilbert damping parameter $\alpha = 0.14$. (b) *R* and *X* data measured at $f = 2$ GHz while the solid lines are the analytical fits to the data.